\RequirePackage{amsmath}
\documentclass[runningheads]{llncs}

\usepackage{amsfonts,bm}
\usepackage{mathtools}
\usepackage{graphicx}
\usepackage{float}
\usepackage{hyperref} 
\usepackage{cuted}
\usepackage{lipsum, color}

\usepackage{latexsym,amssymb}
\usepackage{multirow}
\usepackage{url}

\usepackage{amsmath, amsthm,epsfig,bm,booktabs,amssymb}
\usepackage[ruled,vlined]{algorithm2e}
\usepackage{amsmath, amsfonts}
\usepackage{amsmath,latexsym,amssymb,epsf,amsfonts,amsthm}
\usepackage{epsf}
\usepackage{color}
\usepackage{mathrsfs}
\usepackage{epstopdf}
\usepackage{threeparttable}
\usepackage{graphicx}
\usepackage{tabularx}
\usepackage[all]{xy}

\usepackage{tikz}
\usetikzlibrary{positioning}
\usetikzlibrary{shapes,arrows,arrows,positioning,fit}

\newlength\plotwidth
\hfuzz=\maxdimen
\def \R {{\mathbb R}}

\newcommand{\be}{\begin{equation}}
\newcommand{\ee}{\end{equation}}
\newcommand{\ba}{\begin{eqnarray}}
\newcommand{\ea}{\end{eqnarray}}
\newcommand{\bi}{\begin{itemize}}
\newcommand{\ei}{\end{itemize}}

\newcommand{\comments}[1]{}

\begin{document}
\title{\texorpdfstring{Computing Residual Diffusivity by \\Adaptive Basis Learning via \\Super-Resolution Deep Neural Networks}{Lg}}
\titlerunning{Deep Adaptive Basis Learning}
\author{
Jiancheng Lyu
\and
Jack Xin
\and 
Yifeng Yu
}
\institute{
UC Irvine, Irvine, CA 92697, USA.
\email{(jianchel,jack.xin,yifengy)@uci.edu}}
\authorrunning{J. Lyu, J. Xin and Y. Yu}
\date{}
\maketitle

\begin{abstract}
It is expensive to compute residual diffusivity in chaotic in-compressible flows by solving advection-diffusion equation due to the formation of sharp internal layers in the advection dominated regime. Proper orthogonal decomposition (POD) is a classical method to construct a small number of adaptive orthogonal basis vectors for low cost computation based on snapshots of fully resolved solutions at a particular molecular diffusivity $D_{0}^{*}$. The quality of POD basis deteriorates if it is applied to $D_0\ll D_{0}^{*}$. To improve POD, we adapt a super-resolution generative adversarial deep neural network (SRGAN) to train a nonlinear mapping based on snapshot data at two values of $D_{0}^{*}$. The mapping models the sharpening effect on internal layers as $D_0$ becomes smaller. We show through numerical experiments that after applying such a mapping to snapshots, the prediction accuracy of residual diffusivity improves considerably that of the standard POD. 
\end{abstract}






\section{Introduction}
\setcounter{equation}{0}
Diffusion enhancement in fluid advection has been studied for nearly a century, 
since the pioneering work of Taylor \cite{T_21}. A fundamental problem is to characterize and quantify the large scale effective diffusion (denoted by $D^E$) in fluid flows containing complex and turbulent streamlines. Much progress has been made based on the passive scalar model \cite{MK_99}:
\be\label{eqn1}
T_t+\left(\boldsymbol{v}\cdot D \right) T = D_0 \; \Delta T,
\ee
where $T$ is a scalar function (e.g. temperature or concentration), $D_0 > 0$ is a constant (the so called molecular diffusivity), $\boldsymbol{v}\left(\boldsymbol{x},t\right)$ is a prescribed incompressible velocity field, $D$ and $\Delta $ are the spatial gradient and Laplacian operators.
  
When the flow is steady, periodic and two dimensional, precise asymptotics of $D^E$ are known. 
A prototypical example is the steady cellular flow \cite{CG95,FP94}, 
 $\boldsymbol{v}= (-H_y, H_x)$, $H=\sin x \sin y$, see also \cite{NR_07,XY_13,XY_14a} for 
its application in effective speeds of front propagation. The asymptotics of 
the effective diffusion along any unit direction in the cellular flow 
obeys the square root law in the advection dominated regime: 
$D^{E}=O(\sqrt{D_0})\gg D_0$ as $D_0 \downarrow 0$, \cite{FP94,H03}. 
This is due to the ordered streamlines of the steady cellular flows where 
enhanced transport occurs along saddle to saddle connections and 
a diffusing particle escapes closed streamlines by hopping from cell to cell. 
However, if the streamlines are fully {\bf chaotic} (well-mixed), the enhancement 
can follow a very different law. The simplest such example is the 
time periodic cellular flow:
\be\label{tdcell}
\boldsymbol{v}=(\cos (y),\cos (x) ) +  \theta \, \cos (t)\;(\sin (y),\sin (x)),
\quad
\theta \in (0,1].
\ee 
The first term of (\ref{tdcell}) is a steady cellular flow with a $\pi/4$ 
rotation, and the second term is a time periodic perturbation that introduces 
an increasing amount of disorder in the flow trajectories as $\theta$ becomes larger. At $\theta =1$, the flow (\ref{tdcell}) is fully chaotic
\cite{ZCX_15},  
and has served as a model of 
chaotic advection for Rayleigh-B\'enard experiment \cite{CW91}.
Numerical simulations \cite{BCVV95,LXY_17b,WXZ_18} suggest that at $\theta = 1$, 
the effective diffusivity along the $x$-axis, $D^E_{11} = O(1)$ as $D_0 \downarrow 0$, 
the so called {\it residual diffusivity} phenomenon emerges.  As $D_0 \downarrow 0$, the solutions develop 
sharp gradients, and render accurate computation costly. 
The effective diffusivity tensor is given by \cite{BLP2011}:
\be
D^{E}_{ij} = D_0 \left ( \delta_{ij} +\left \langle  D w_i \cdot D w_j \right\rangle \right), \label{De1}
\ee
where $w=(w_1,w_2)$ is the unique mean zero space-time periodic vector solution of the 
{\it cell problem} \cite{BLP2011}:
\be
w_t + (\boldsymbol{v}\cdot D w) - D_0 \Delta w = - \boldsymbol{v}, \label{De2}
\ee
and the bracket in (\ref{De1}) denotes space-time average over the periods. 
The second term of (\ref{De1}) is a positive definite correction to $D_0\, \delta_{ij}$.

The solution of (\ref{De2}) subject to periodic boundary condition in space 
can be computed accurately by spectral method in Fourier basis. The drawback is that 
the number of Fourier modes grow rapidly as $D_0 \downarrow 0$. In \cite{LXY_17b}, 
adaptive orthogonal basis vectors are constructed from snapshots of spectral solutions 
to handle the near singular solutions of (\ref{De2})
at small $D_0$. First, $w$ is computed by spectral method with  
cellular flow $\boldsymbol{v}$ represented compactly in Fourier space. 
Truncating the Fourier expansion leads to an approximate 
system of ordinary differential equations (ODEs). The time periodic solution is constructed 
as the unique fixed point of the Poincar\'e map of the ODE's time $2\pi$ flow, bypassing 
long time simulation. The snapshots of solutions in the time interval $[0,2\pi]$ are 
saved into columns of a matrix $W$. The adaptive basis vectors are left singular eigenvectors 
corresponding to the top singular values of $W$. This is the training process for adaptive basis, and is done at a few sampled $D_0$ or $\theta$ values. The equation (\ref{De2}) 
at other $D_0$ or $\theta \in (0,1]$ are solved 
in terms of the adaptive basis trained at the closest sample $D_0$ or $\theta$ value. Then 
formula (\ref{De1}) is used to calculate $D^{E}$. The number of adaptive basis functions 
is found to be much less than that of Fourier basis by several orders of magnitude. 
The relative error of the adaptive solution from a resolved spectral solution is under 6.5 \% 
when testing at $D_0=10^{-5}$ and training at $D_0=10^{-4}$. 

The above procedure to generate adaptive basis vectors by taking snapshots of solutions and performing singular value decomposition (SVD) is 
the so called reduced order modeling \cite{Sir_87,Quart_14} and   
known as proper orthogonal decomposition (POD) in the fluid dynamics literature \cite{Lum_67,HL_98}. 
The residual diffusivity problem here offers an ideal testing ground for 
the evaluation of POD.  The time periodicity of $\boldsymbol{v}$ reduces the 
evolution problem (\ref{De2}) to a Poincar\'e map problem. Snapshots (training data) 
are directly drawn from the time periodic solution which is the attractor of (\ref{De2}). 
These steps help the testing accuracy of the reduced order solutions  
at the out-of-sample $D_0$ or $\theta$ values. However, 
POD method depends on collection of good snapshots data. For example when snapshots are collected at certain $D_0$ and the solutions at a much smaller $D_1 \ll D_0$ are to be computed with reduced basis constructed at $D_0$, the errors can rapidly increase. 
\medskip

In this paper, we study a deep neural network (DNN) approach to alleviate the accuracy loss  of POD and improve the error of reduced basis computation at $D_1$ based on prediction from snapshots at two values  
$D_{0}^{1}$ and $D_{0}^{2}$ (both above $D_1$). The idea is to train a mapping from snapshots (images) at $D_{0}^{1}$ to those at $D_{0}^{2}$ ($D_{0}^{2}< D_{0}^{1}$). The mapping sharpens 
the images similar to what happens to solutions of 
(\ref{eqn1}) as $D_0$ becomes smaller. 
The mapping is applied to 
snapshots at $D_{0}^{2}$ for improving POD  
basis construction. As a proof of concept, we select 
a super-resolution DNN in the form of a generative adversarial network (SRGAN, \cite{LTHCCAATTWS2017}). 
We show that it serves our purpose well through numerical experiments where the mapping is constructed (trained) based on snapshots at $D_{0}^{1}$ and $D_{0}^{2}$, then tested (applied) at $D_1< D_{0}^{2}$. 
\medskip

The paper is organized as follows. 
In section 2, we describe the POD basis construction for (\ref{De1}),  
SRGAN architecture and 
its training objective. In section 3, we present computational results on predicted residual  diffusivity from POD basis with and without 
SRGAN. Concluding remarks are in section 4. 
\medskip

\section{Construction of Adaptive Basis via DNNs}
\setcounter{equation}{0}
\subsection{Learning thinner structures}

Consider spectral method for computing the cell problem below for $D^{E}_{11}$: 
\begin{align}\label{cell1}
w_t+\left(\boldsymbol{v}\cdot\nabla \right ) w- D_0\, \Delta \, w = -v,
\end{align}
where $\boldsymbol{v}=(v,\tilde{v})$, and 
\begin{equation}\label{flowform_1}
\begin{split}
v\left(x,t\right) = \cos\left(x_2\right)+\sin\left(x_2\right)\cos\left(t\right),\\
\tilde{v}\left(x,t\right) = \cos\left(x_1\right)+\sin\left(x_1\right)\cos\left(t\right).
\end{split}
\end{equation}
Let $w_{k}^{N}$ be the $k$-th mode of a $(2N+1)^2$ term Fourier approximation of $w$ on the $[0, 2\pi]^2$ periodic domain \cite{LXY_17b}. Let $v_k$ and $\tilde{v}_{k}$ be the $k$-th Fourier modes of $v$ and $\tilde{v}$ respectively. In view of (\ref{flowform_1}),  
$v_k$ ($\tilde{v}_{k}$) equals zero unless $k=(0,\pm 1)$ ($k=(\pm 1,0)$).
%
The truncated ODE system on $w_{k}^{N}$ is:
\begin{equation}\label{odet_1}
\dfrac{dw^N_{k}}{dt}+D_0\left|k\right|^2w^N_{k}+i\sum_{\left\|k-j\right\|\leq N}\left[\left(k_1-j_1\right)v_{j}\left(t\right)+\left(k_2-j_2\right)\tilde{v}_{j}\left(t\right)\right]w^N_{k-j} = -v_{k}\left(t\right),
\end{equation}
which reads in vector-matrix form:
$d\mathbf{w}/dt = A\left(t\right)\mathbf{w}+\mathbf{v}\left(t\right).$
For a time discretization with $N_t$ grid points on $[0,2\pi]$, let $\left\{\hat{\mathbf{w}}^{*}_n\right\}_{n=0}^{N_t}$ be a numerical time periodic solution to \eqref{odet_1} for $D_0=D_0^*$, a value where snapshot data are collected. Such a solution can be obtained by finding the unique fixed point of the  Poincar\'e map \cite{LXY_17b}. Define the solution matrix of size $(2N+1)^2 \times N_t$:
\begin{align}\label{smt}
W = \left[\hat{\mathbf{w}}^{*}_0\quad \hat{\mathbf{w}}^{*}_1\quad \dots\quad \hat{\mathbf{w}}^{*}_{N_t}\right],
\end{align}
and compute the SVD factorization $W = U\, \Sigma\, V^{T}$. Then one extracts columns $\mathbf{u}_j$ ($j=1,\cdots,m$) 
of the matrix $U$ corresponding to the largest $m \ll O(N^2)$ singular values, to form the adaptive orthogonal 
basis vectors and the matrix: 
$U_m = \left[\mathbf{u}_1\quad \mathbf{u}_2\quad \dots\quad \mathbf{u}_m\right] $. 
This is the end of basis training at a sampled value $D^{*}_{0}$. 
At $D_0 \not = D^{*}_{0}$, project $\mathbf{w}\left(t\right)$ to the span of column vectors of $U_m$ or seek a vector of the form $U_m\,  \mathbf{a}\left(t\right)$, 
where $\mathbf{a} (t) \in \R^m$ satisfies the ODE system in a much lower dimension (bar is complex conjugate, $T$ is transpose):
\begin{align}\label{odea}
\dfrac{d\mathbf{a}}{dt} = \bar{U}_m^TA\left(t\right)U_m\mathbf{a}+\bar{U}_m^T\mathbf{v}\left(t\right).
\end{align}
Finding the time periodic solution to a time discrete version of  (\ref{odea}) via Poincar\'e map to compute $D^E$ by (\ref{De1}) in Fourier space, we completed the reduced order modeling.   
\medskip

\begin{figure}[ht]
\centering
\begin{tabular}{c}
\includegraphics[width=\plotwidth]{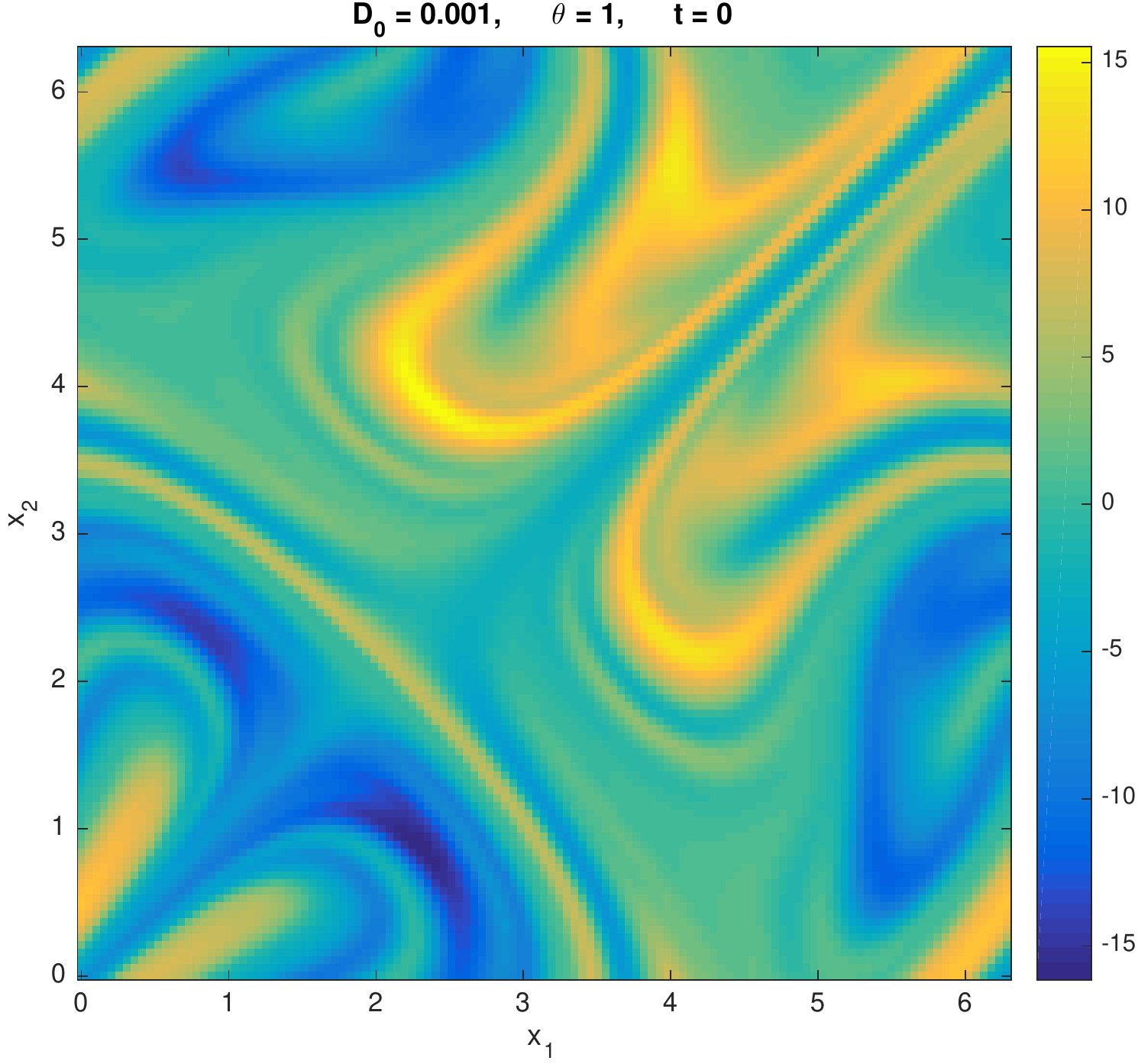}
\includegraphics[width=\plotwidth]{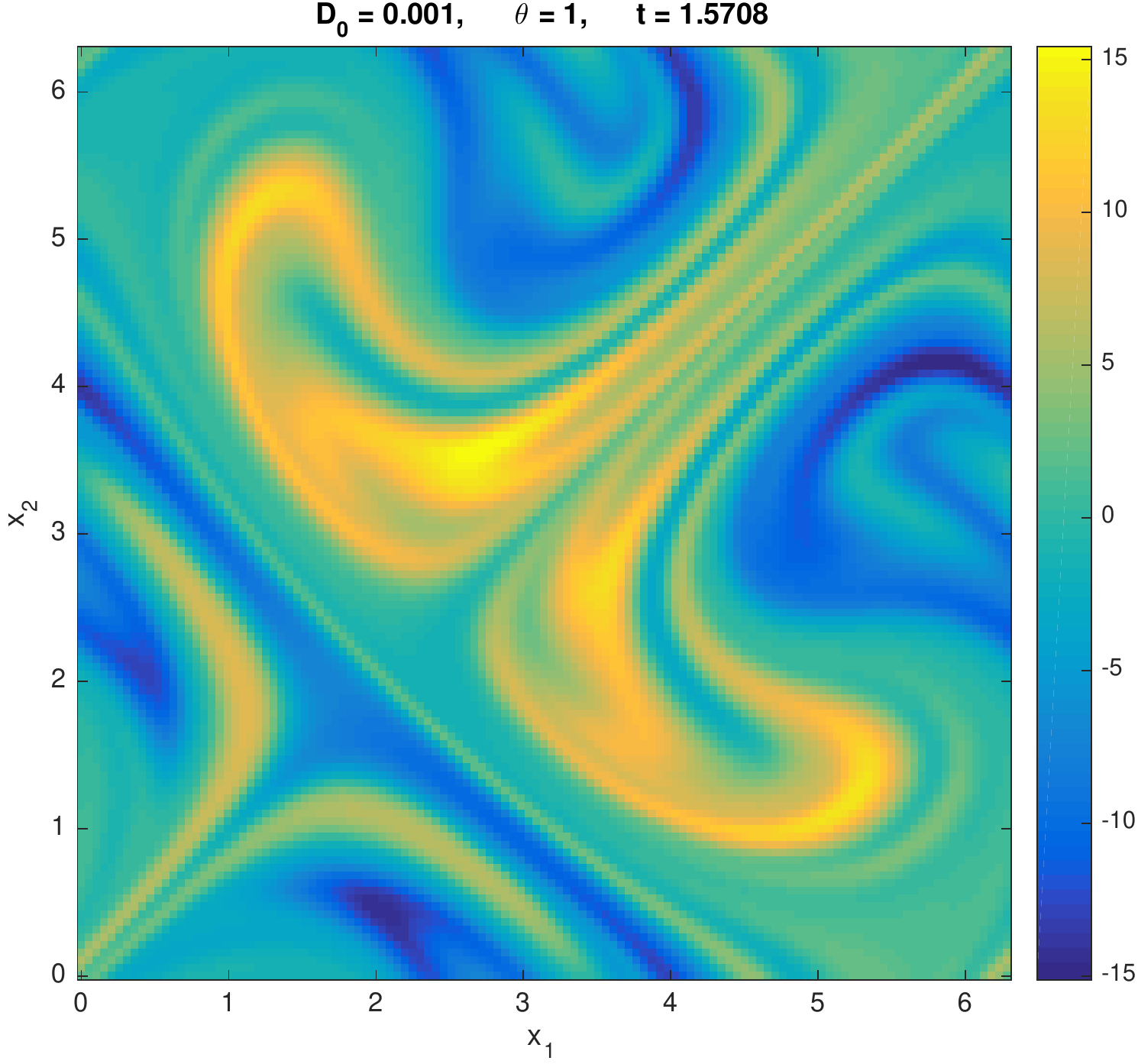}\\
\includegraphics[width=\plotwidth]{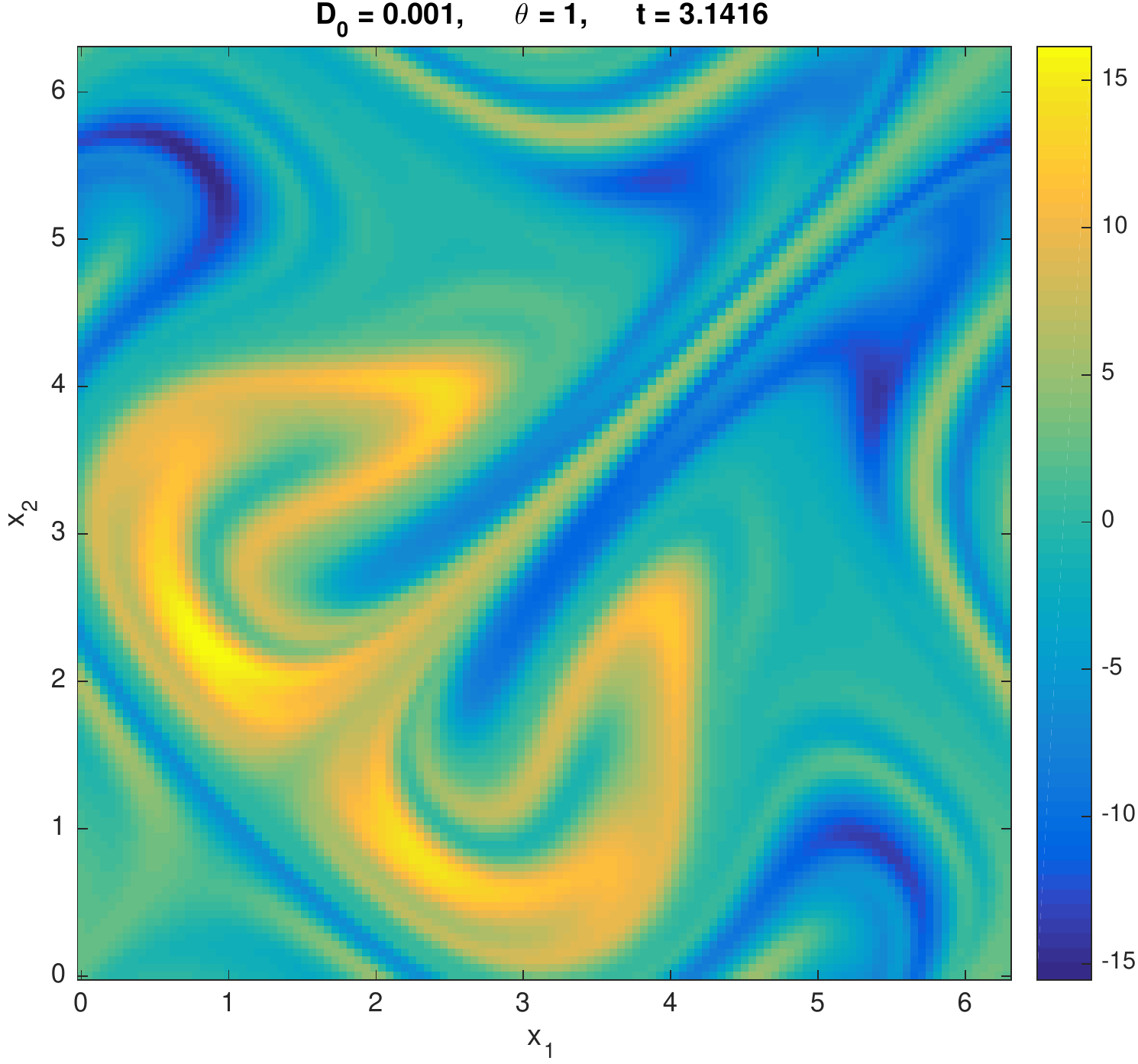}
\includegraphics[width=\plotwidth]{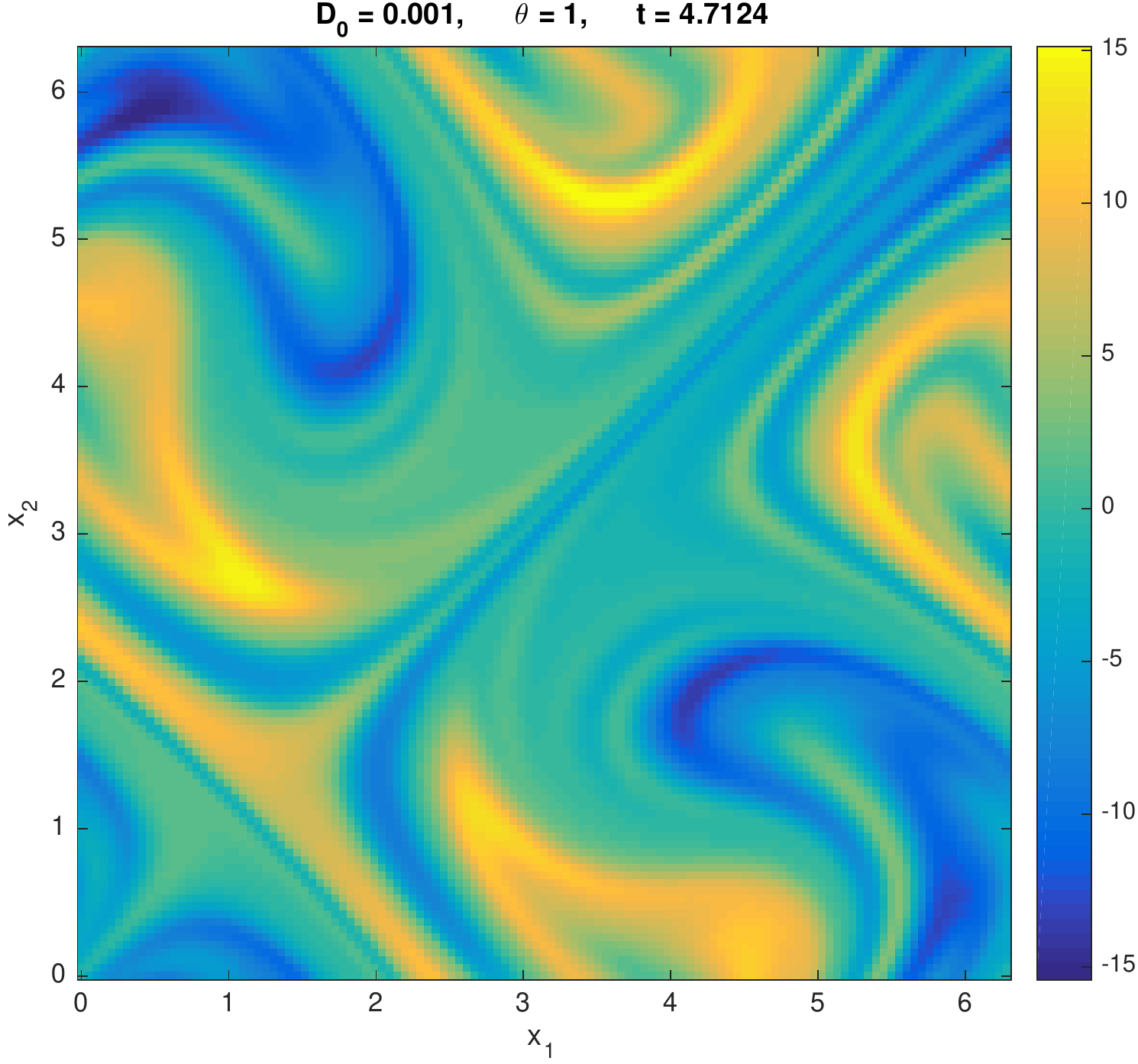}
\end{tabular}
\caption{Sampled snapshots of \eqref{cell1} at $D_0 = 10^{-3}$
with layered structures.}\label{snapshot-3}
\end{figure}

\begin{figure}[ht]
\centering
\begin{tabular}{c}
\includegraphics[width=\plotwidth]{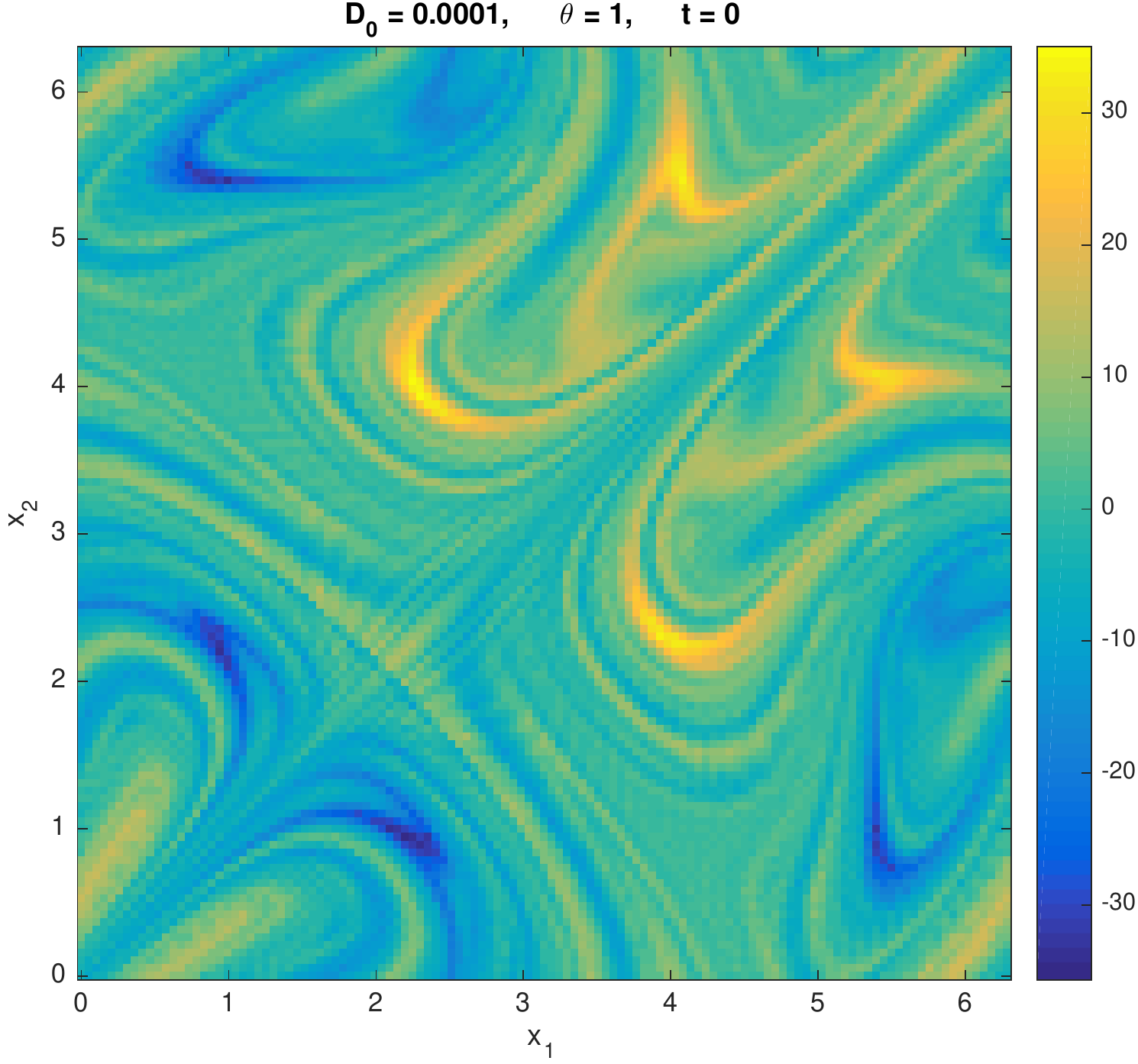}
\includegraphics[width=\plotwidth]{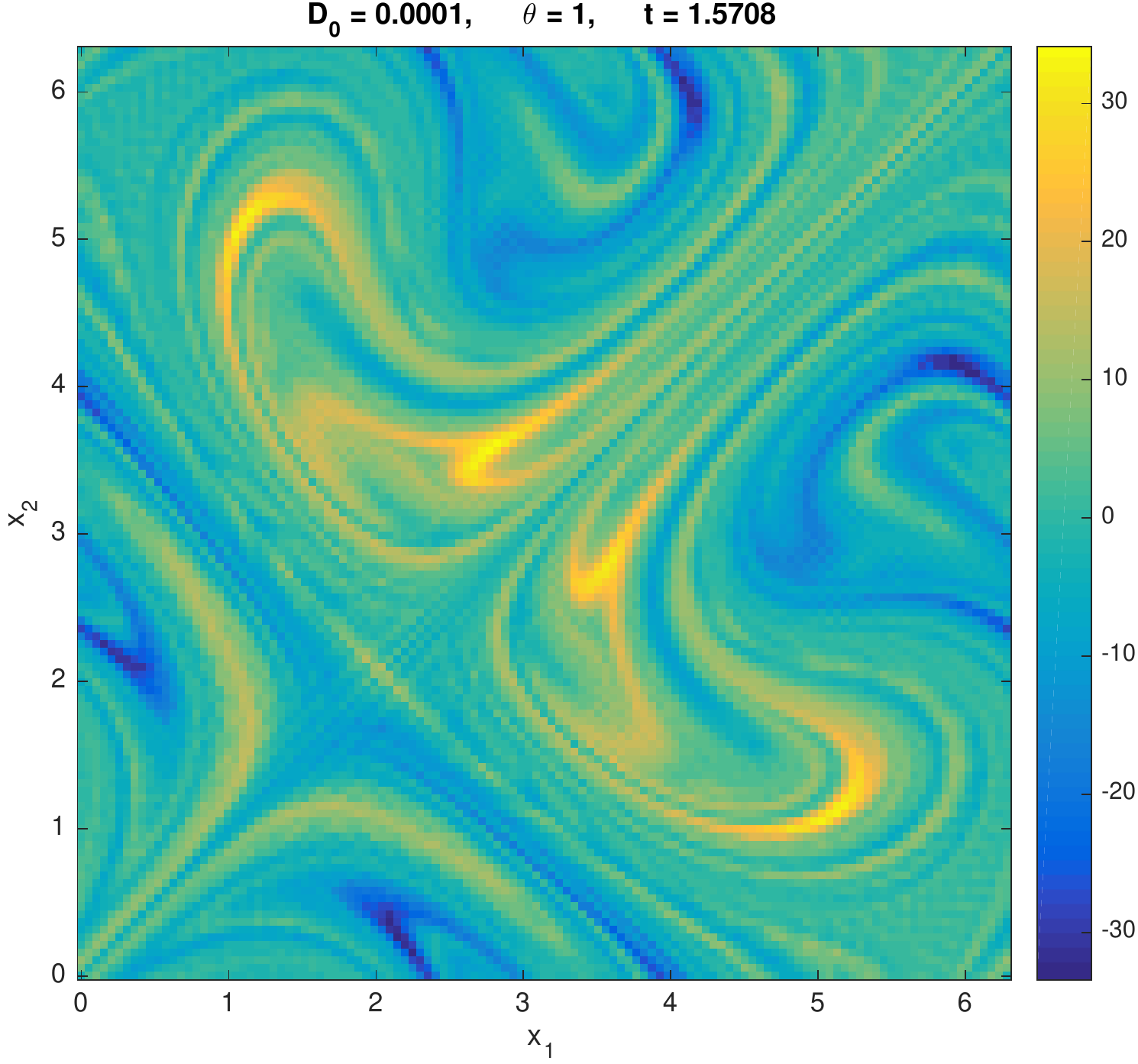}\\
\includegraphics[width=\plotwidth]{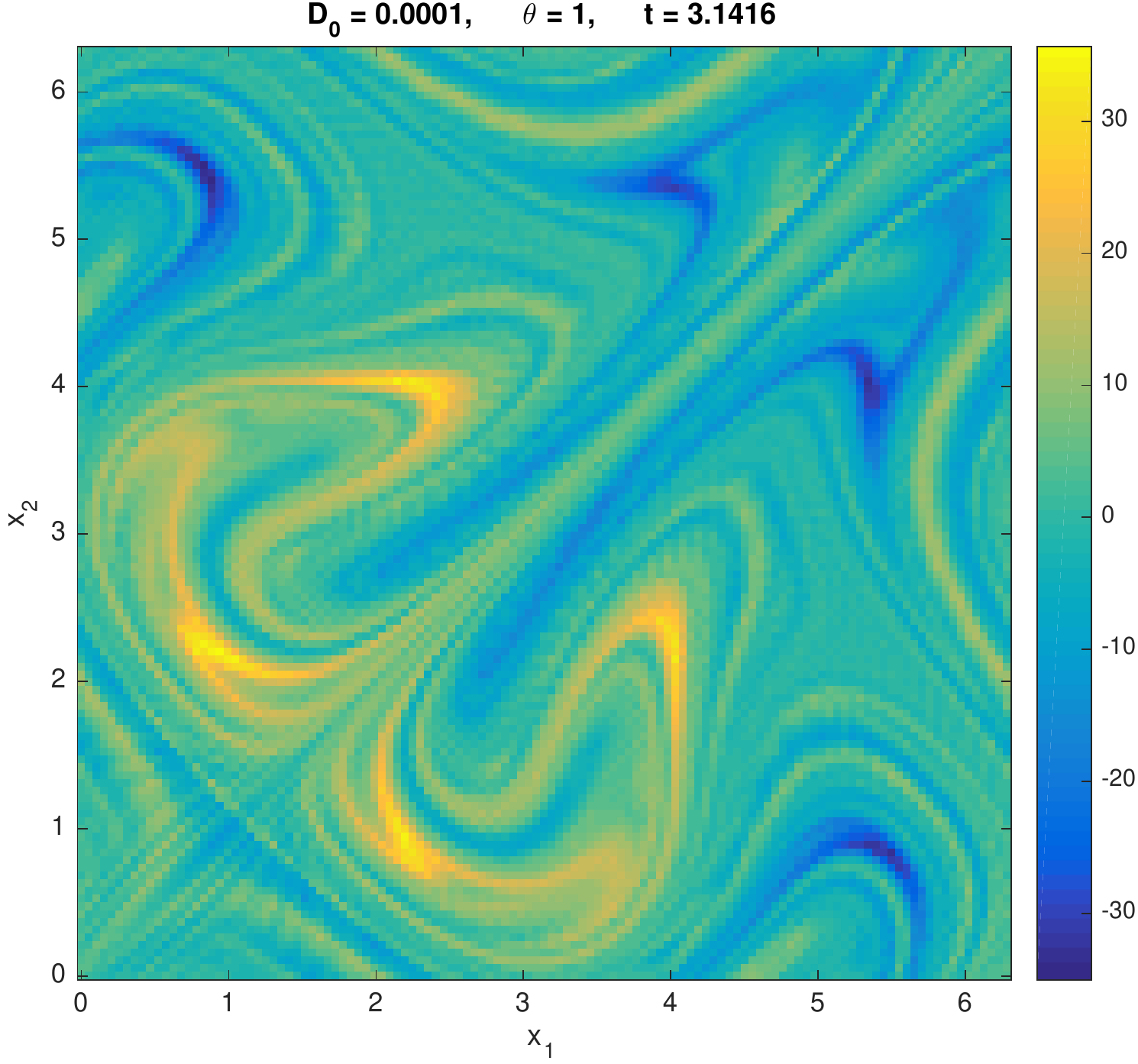}
\includegraphics[width=\plotwidth]{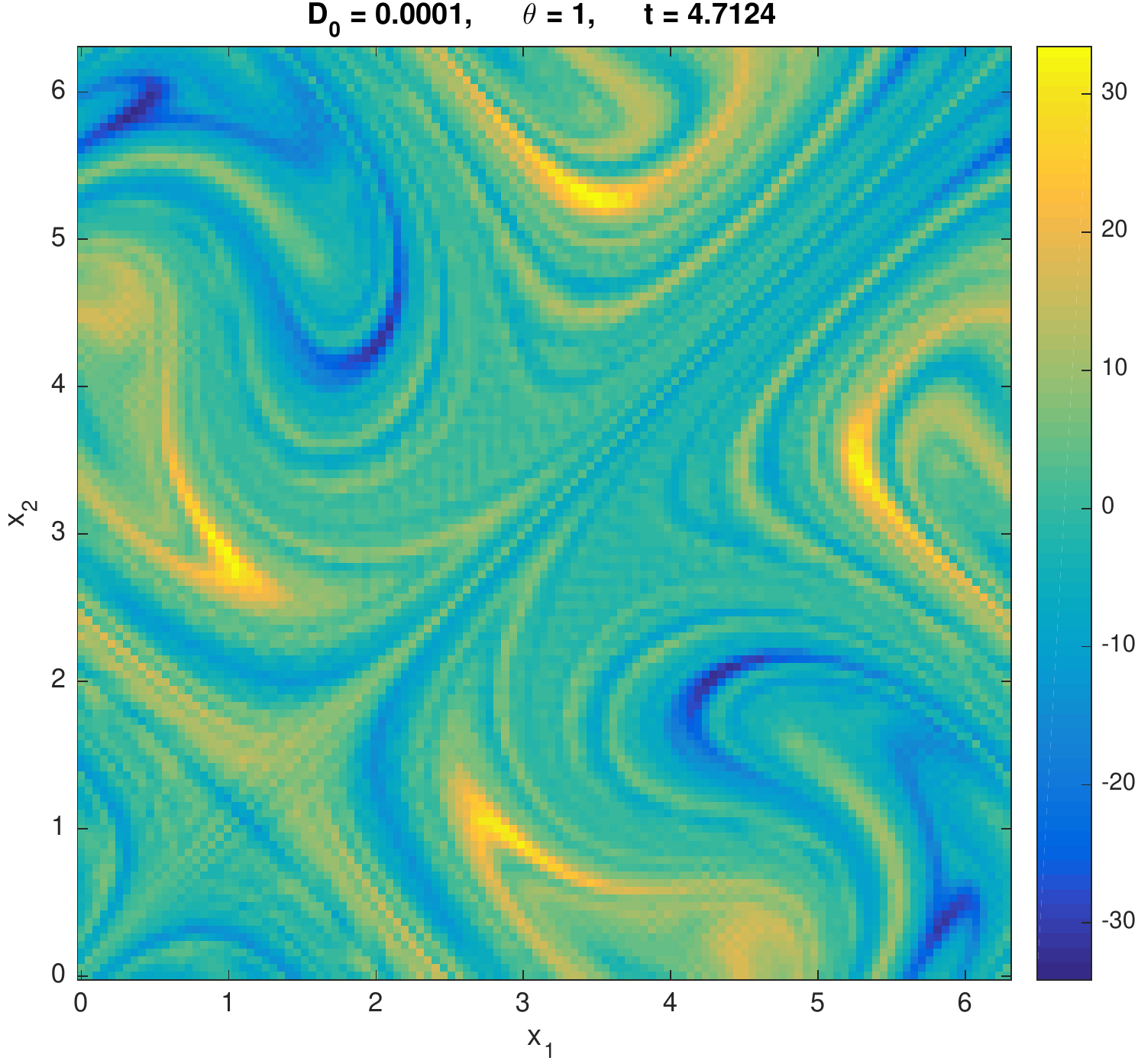}
\end{tabular}
\caption{Sampled snapshots of  \eqref{cell1} at $D_0 = 10^{-4}$, formation of thinner layers.}\label{snapshot-4}
\end{figure}

The inverse Fourier transform of $W$ gives the snapshot matrix in the physical domain:
\begin{align}\label{smt2}
S_{p} = \left[\mathbf{I}^{*}_0\quad \mathbf{I}^{*}_1\quad \dots\quad \mathbf{I}^{*}_{N_t}\right]
\end{align}
where each column vector (snapshot) is an image after reshaping into a square matrix. 
The $S_p$ is convenient for visualization and drawing a connection with image processing. %
%
Figs. \ref{snapshot-3}-\ref{snapshot-4} illustrate that internal layers in the physical domain snapshots (column vectors of $S_p$)  emerge and get thinner as $D_0$ becomes smaller. For a better prediction of the nearly singular solutions at $D_1$ much smaller than $D_0^*$, it is helpful if the adaptive basis learned at $D_0^*$ encodes certain thinner layered structures. Particularly, given the solution matrix $W$ at $D_0^*$ as \eqref{smt}, we look for a map $\mathcal{M}$ such that the physical domain snapshots (inverse Fourier transform) of $\mathcal{M}\left(W\right)$ have sharpened internal layers.

Suppose $D_0^1 > D_0^2$ are two values  $\geq  D_{0}^{*}$, and $W^i$ is the Fourier domain solution matrix at $D_0^i$ for $i = 1, 2$. Let $\mathcal{F}$ be column-wise Fourier transform on matrices, then columns of $\mathcal{F}^{-1}\left(W^i\right)$ are physical domain snapshots of $W^i$. When the $W^i$'s are available, we use DNN to train a map $\mathcal{T}$ for the following regression problem: 
\begin{equation}\label{regress}
 \mathcal{T}: \mathcal{F}^{-1}\left(W^1\right) \to \mathcal{F}^{-1}\left(W^2\right).
\end{equation}
Since $D_0^2 < D_0^1$,  $\mathcal{F}^{-1}\left(W^2\right)$ has thinner layered structures than $\mathcal{F}^{-1}\left(W^1\right)$. By solving the regression problem (\ref{regress}) 
via DNN, our goal is that the  $\mathcal{T}\left(\mathcal{F}^{-1}\left(W^1\right)\right)$ inherits the image sharpening capability. Then  $\mathcal{T}$ can be applied to solution matrix $W^*$ at $D_{0}^{*} \leq D_{0}^{2}$ and $\mathcal{T}\left(\mathcal{F}^{-1}\left(W^*\right)\right)$ is expected to have thinner structures for better prediction of residual diffusivity. Finally, the adaptive basis with thinner structures will 
be obtained from SVD of 
\begin{align*}
    \mathcal{M}\left(W^*\right) = \mathcal{F}\left(\mathcal{T}\left(\mathcal{F}^{-1}\left(W^*\right)\right)\right).
\end{align*}

\subsection{Adversarial network}
We opt for the super-resolution generative adversarial network (SRGAN) \cite{LTHCCAATTWS2017} to train the map $\mathcal{T}$. As a generative adversarial network (GAN), SRGAN consists of a generator network $G$ and a discriminator network $D$. The two networks compete in a way that $D$ is trained to distinguish the real high-resolution (HR) images and those generated from low-resolution (LR) images, while $G$ is trained to create fake HR images from LR images to fool $D$. We train the SRGAN with $\mathcal{F}^{-1}\left(W^1\right)$ as input data and $\mathcal{F}^{-1}\left(W^2\right)$ as target data so that the generator $G$ learn to generate thinner structures when it is fed with $\mathcal{F}^{-1}\left(W\right)$. In this approach, we realize $\mathcal{T}$ through a trained $G$.
%

\begin{figure}[ht]
\centering
\begin{tabular}{cc}
\multicolumn{2}{c}{\includegraphics[width=.8\textwidth]{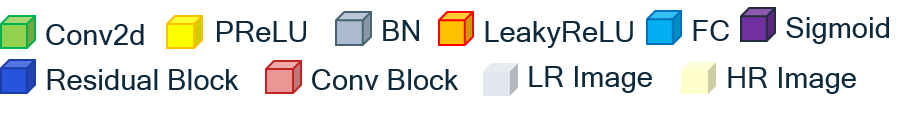}}\\
\multicolumn{2}{c}{\includegraphics[width=.85\textwidth]{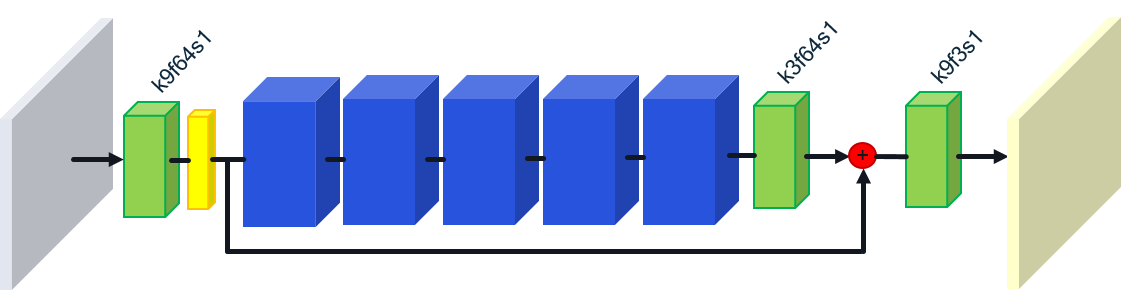}}\\
\multicolumn{2}{c}{(a)}\\
\multicolumn{2}{c}{\includegraphics[width=.85\textwidth]{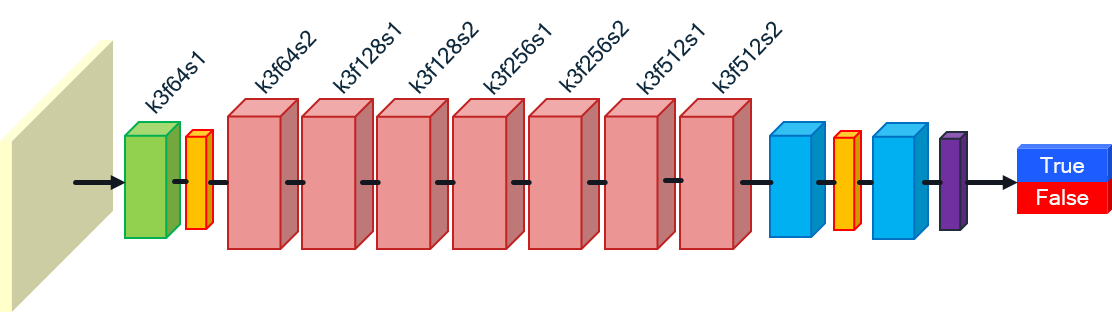}}\\
\multicolumn{2}{c}{(b)}\\
\includegraphics[width=.35\textwidth]{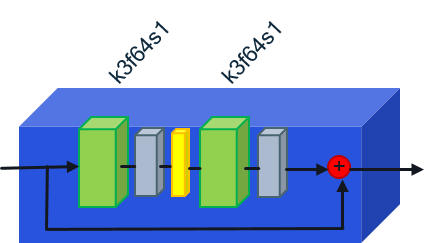} &
\includegraphics[width=.25\textwidth]{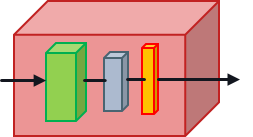}\\
(c) & (d)
\end{tabular}
\caption{Architecture of the generator and discriminator networks. (a) The generator network. (b) The discriminator network. (c) Residual block in the generator network. (d) Convolutional block in the discriminator network. We use a simple notation to indicate the Conv2d layer. For example, k9f64s1 indicates a convolutional layer with kernel size $9$, number of filters $64$ and stride $1$.}\label{network}
\end{figure}

The network architecture is shown in Figure \ref{network}. The generator network $G$ starts with a convolutional block with kernel size $9\times9$, followed by a few residual blocks. Here a convolutional block consists of a convolutional layer and a PReLU layer, a residual block is a convolutional block with kernel size $3\times3$ followed by a convolutional layer of the same kernel size and a shortcut from the input to output. There are two more convolutional layers with kernel size $3\times3$ and $9\times9$ after the residual blocks at the end of the network. The number of filters in all convolutional blocks are the same except for the last one. Note that we remove the two upscale layers in \cite{LTHCCAATTWS2017} since the snapshot sizes of $\mathcal{F}^{-1}\left(W^1\right)$ and $\mathcal{F}^{-1}\left(W^2\right)$ are the same. 
\medskip

The discriminator network $D$ is defined by the architectural guidelines summarized in \cite{RMC2016}, see Figure \ref{network}. It has eight convolutional blocks with PReLU layers replaced by LeakyReLU layers with slope  parameter $\alpha = 0.2$. Moreover, there is a batch normalization layer before each LeakyReLU in the convolutional blocks. The kernel size is $3\times3$ in all convolutional blocks and the number of filters is doubled in the 3rd, 5th and 7th block. Those blocks are followed by a fully connected layer, a LeakyReLU layer and one more fully connected layer. Finally the feature map is fed in a sigmoid layer which gives the probability of real HR image and the reconstructed image.

As a binary classifier, the discriminator network is equipped with the cross entropy loss. Let us focus on the loss function of the generator network.
Suppose $\mathcal{F}^{-1}\left(W^1\right)$ and $\mathcal{F}^{-1}\left(W^2\right)$ are real matrices of dimension $\left(2N+1\right)^2\times N_t$, and the columns of $\mathcal{F}^{-1}\left(W^1\right)$ and $\mathcal{F}^{-1}\left(W^2\right)$ are $x_i$ and $y_i$, $i = 1, 2, \dots, N_t$. Following the formulation in \cite{LTHCCAATTWS2017}, we define the loss function of the generator network as
\begin{align}\label{lossfun}
    l\left(G\right) = l_{MSE}\left(G\right)+10^{-2}\,l_{VGG}\left(G\right)+10^{-3}\, l_{Gen}\left(G\right).
\end{align}
In \eqref{lossfun}, $l_{MSE}$ is the pixel-wise {\bf MSE loss} defined as the sum of the squares of error at each pixel,
\begin{align*}
    l_{MSE}\left(G\right) = \sum_{i=1}^{N_t}\left\|y_i-G\left(x_i\right)\right\|_2^2.
\end{align*}
The $l_{VGG}$ is the {\bf VGG loss} based on layers of the pre-trained VGG-19 network \cite{vgg_15}. Let $\phi$ be a feature map of VGG-19 and $s_\phi$ be its size, then the VGG loss is the average of squares of Euclidean distances between the feature representations of $y_i$ and $G\left(x_i\right)$
\begin{align*}
    l_{VGG}\left(G\right) = s_{\phi}^{-1}\sum_{i=1}^{N_t}\left\|\phi\left(y_i\right)-\phi\left(G\left(x_i\right)\right)\right\|_2^2.
\end{align*}
The generator network is expected to fool the discriminator network, so \eqref{lossfun} contains $l_{Gen}$ called {\bf generative loss}. The $l_{Gen}$ is defined based on the cross-entropy loss of the discriminator network
\begin{align}\label{gloss}
    \sum_{i=1}^{N_t}\log\left[1-D\left(G\left(x_i\right)\right)\right], 
\end{align}
where $D\left(G\left(x_i\right)\right)$ means the binary classification result of the reconstructed HR image by the generator network $G$. 
In practice, 
we define
\begin{align*}
   l_{Gen}\left(G\right) = \sum_{i=1}^{N_t}-\log D\left(G\left(x_i\right)\right)
\end{align*}
for better gradient behavior.

\section{Experimental results of adaptive basis from SRGAN}

Let $D_0^1 = 10^{-2}$, $D_0^2 = 10^{-3}$. We solved for both $W^1$ and $W^2$ via spectral method with $N = 50$ and $N_t = 1500$, then train SRGAN with input data $\mathcal{F}^{-1}\left(W^1\right)$ and target data $\mathcal{F}^{-1}\left(W^2\right)$. The training of SRGAN includes two stages: 1) We train the generator $G$ for $50$ epochs to get a pretrained model; 2) We train the entire SRGAN for $200$ epochs. Adam and SGD optimizers are applied to training of $G$ and $D$ respectively. We set batch size to be $32$ and learning rate to be $10^{-4}$ for both optimizers. The training was carried out on a desktop with Nvidia graphics cards GTX 1080 Ti. We set $D_0^* = D_0^2$ in the following experiments.
\medskip

\begin{figure}[ht]
\centering
\begin{tabular}{c}
\includegraphics[width=\plotwidth]{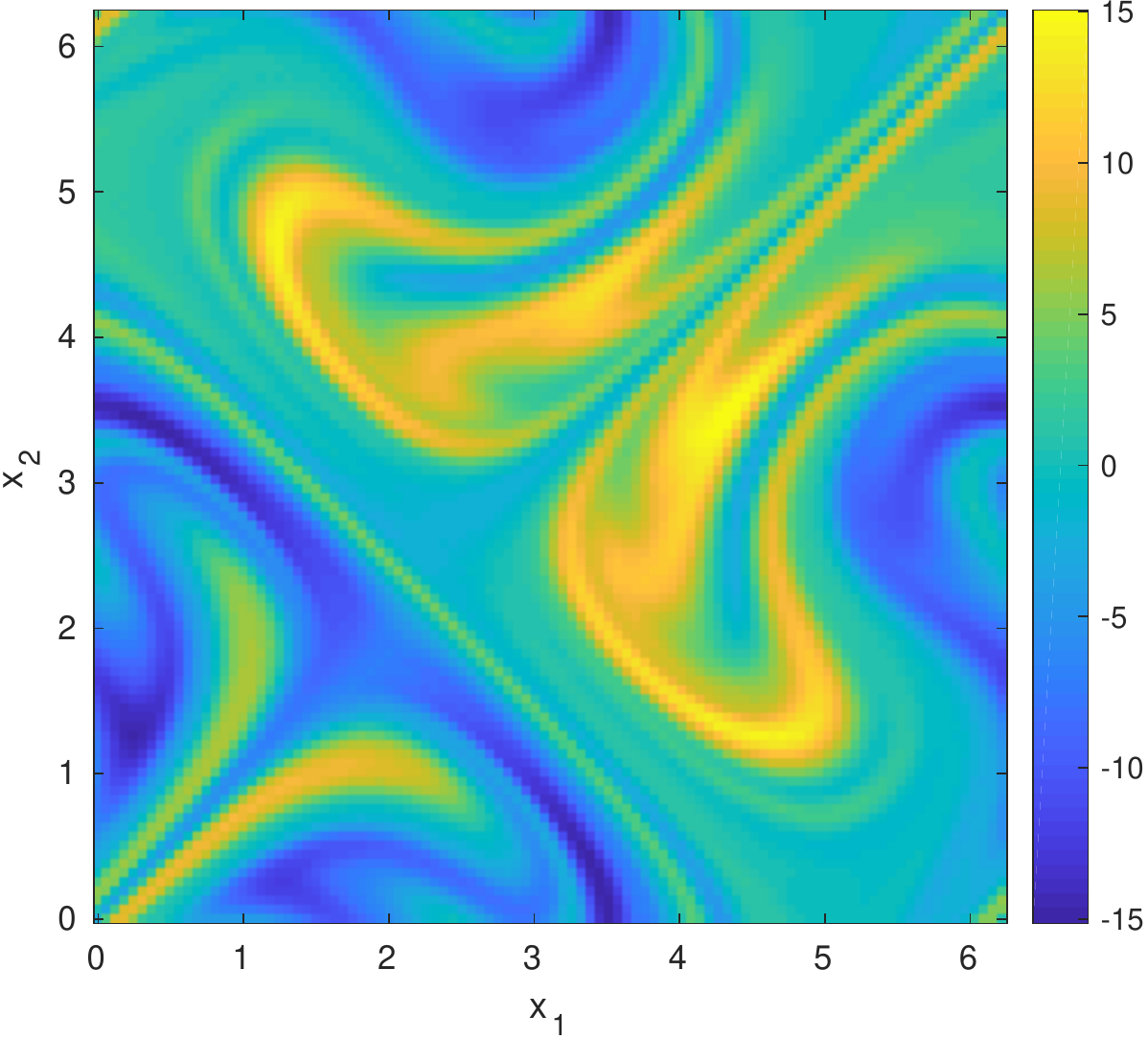}
\includegraphics[width=\plotwidth]{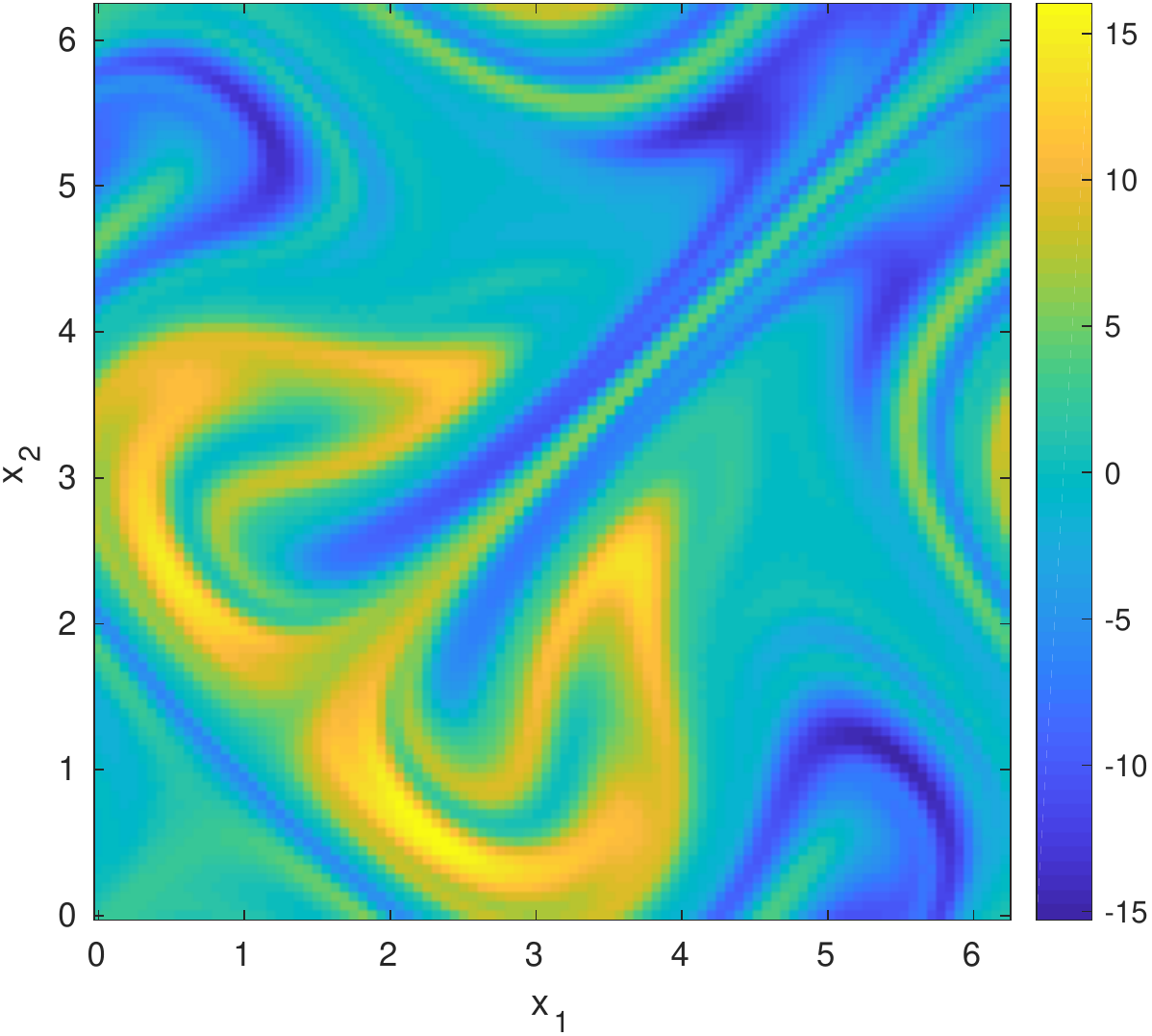}\\
\includegraphics[width=\plotwidth]{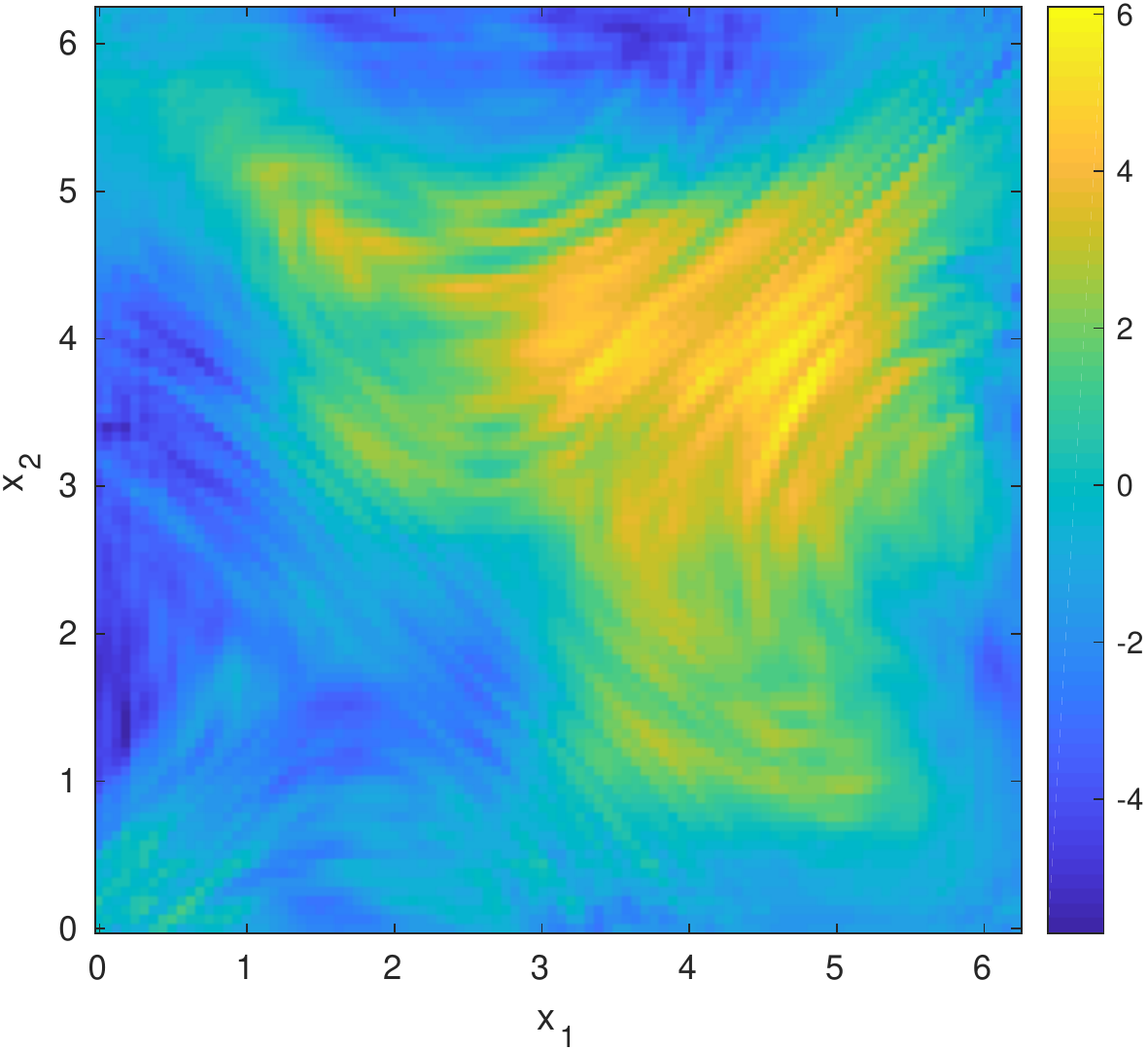}
\includegraphics[width=\plotwidth]{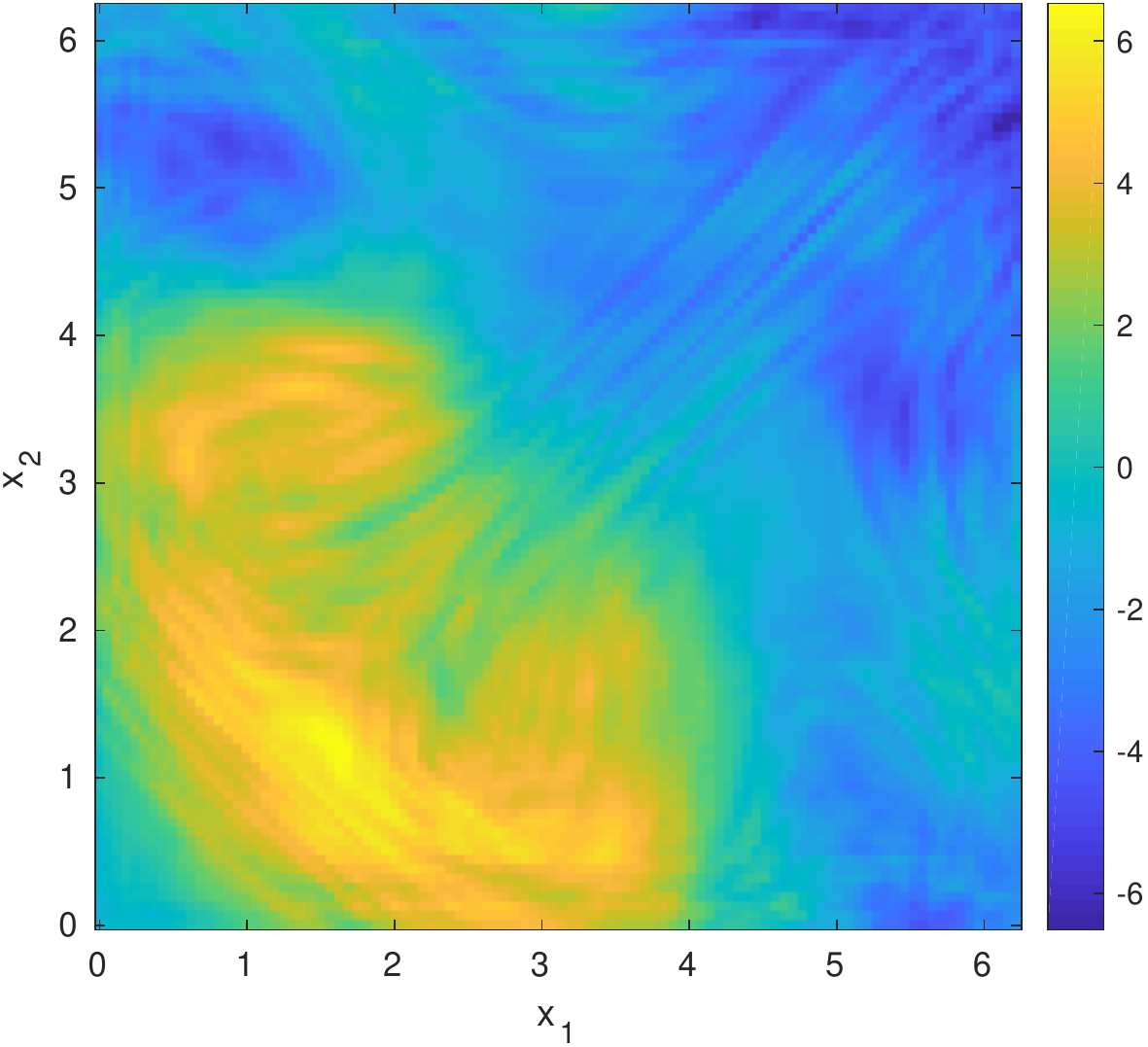}
\end{tabular}
\caption{Input (top) and SRGAN output (bottom) at $D_0^2 = 10^{-3}$.}\label{srganslides}
\end{figure}

Fig. \ref{srganslides} shows two time slices of the input $\mathcal{F}^{-1}\left(W^2\right)$ (top) with  $G\left(\mathcal{F}^{-1}\left(W^2\right)\right)$ (bottom) at $D=D_{0}^{2}$. Columnwise, it can be seen that thinner layers are created by the network $G$. Due to identical dimension constraint 
of the input and output images, the up-scaling layers in SRGAN \cite{LTHCCAATTWS2017} have been removed. This adaptation lowers the 
fidelity of the generated images, as we see in each column of Fig. 
\ref{srganslides}.
However, the SRGAN  generated snapshots are only used to construct reduced basis. The fact that sharp layers are generated by SRGAN training is more important for our task of computing residual diffusivity. 

Set $D_0^1 = 10^{-2}$ and $D_0^2 = 10^{-3}$, the comparison of predictions of $D^E_{11}$ by SVD and SRGAN assisted SVD is shown in Table \ref{srgan1e-2}. The number of adaptive basis is $m= 100$ for both methods.

\begin{figure}[ht]
\centering
\begin{tabular}{c}
\includegraphics[width=\plotwidth]{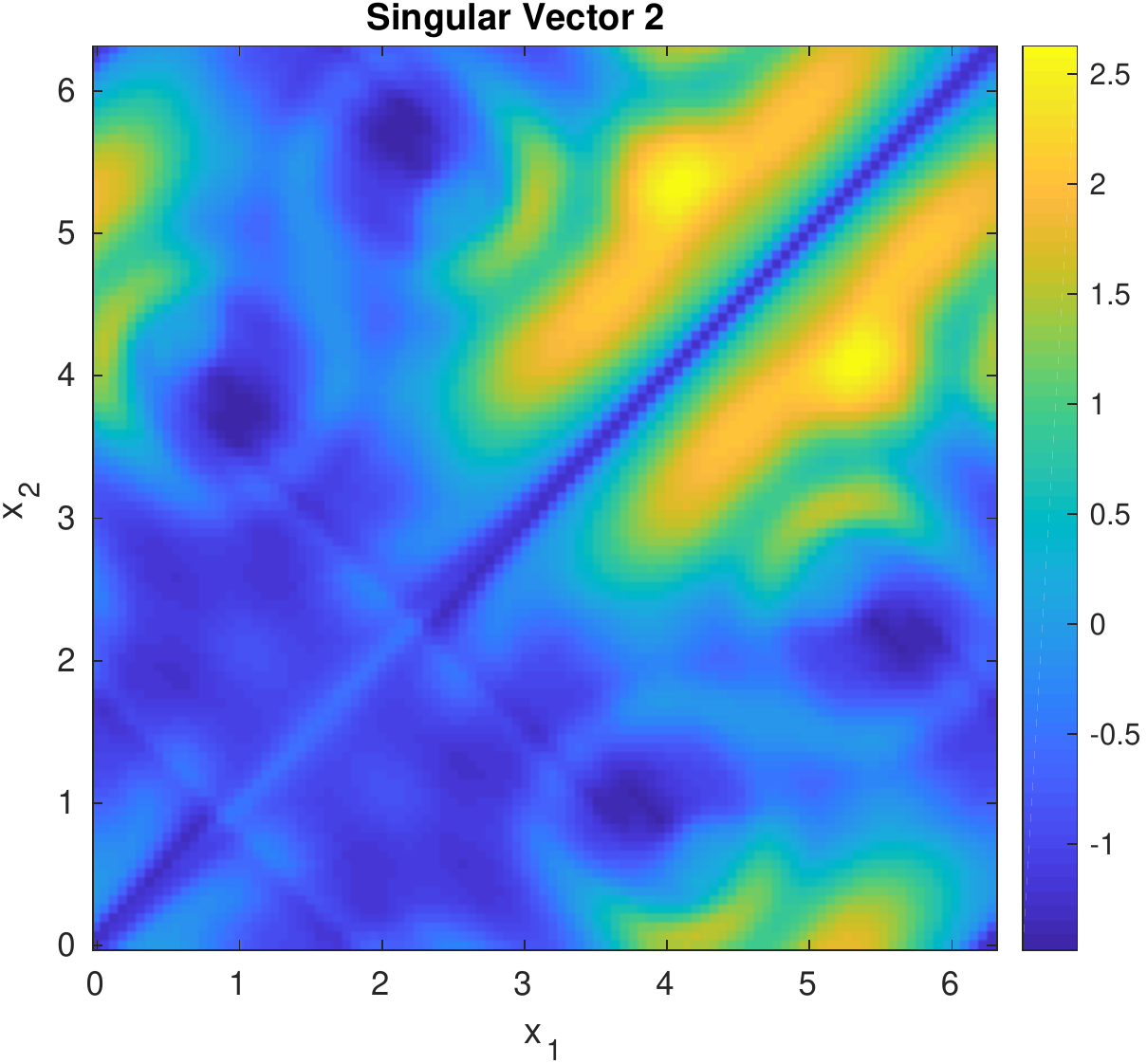}
\includegraphics[width=\plotwidth]{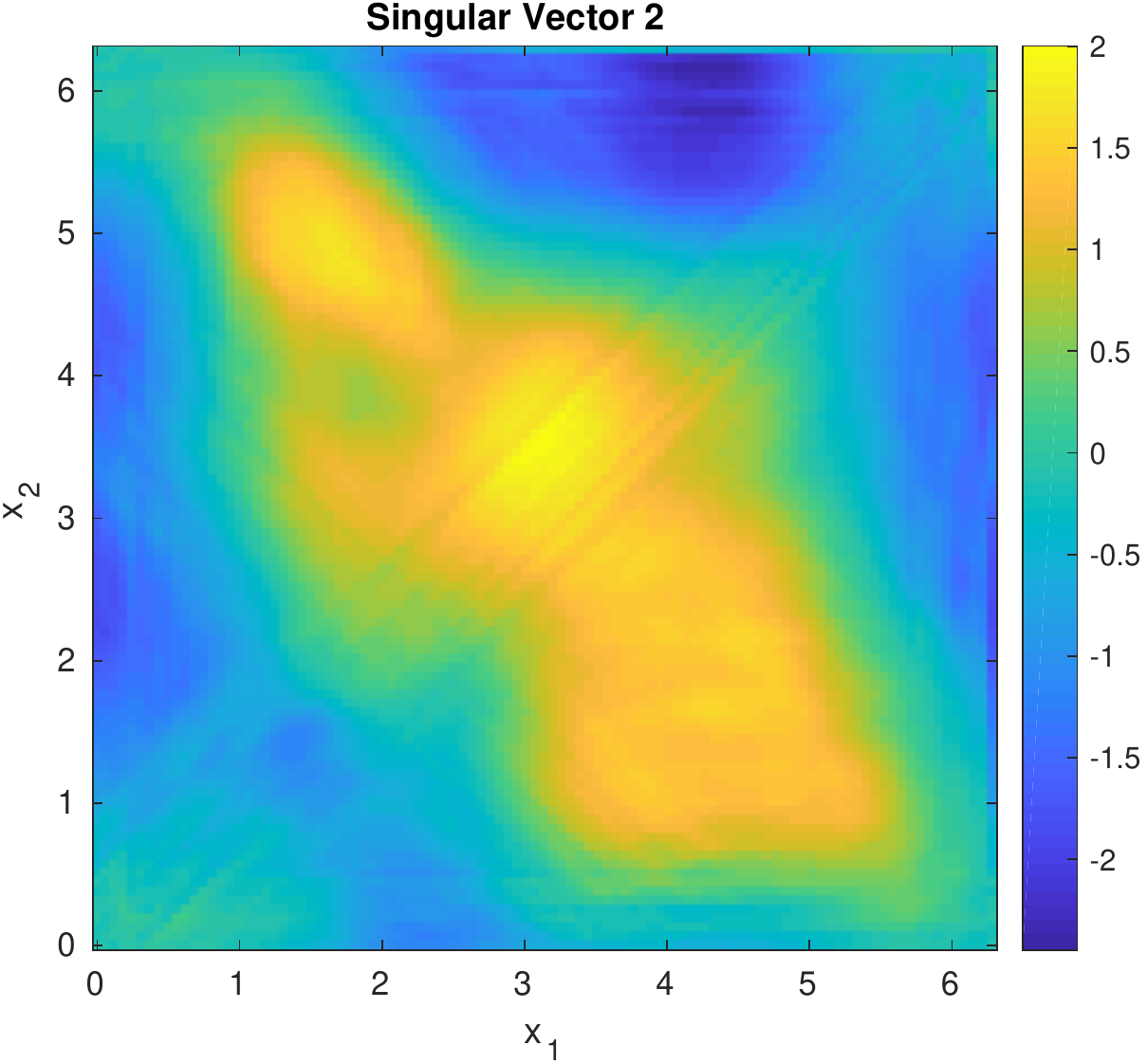}\\
\includegraphics[width=\plotwidth]{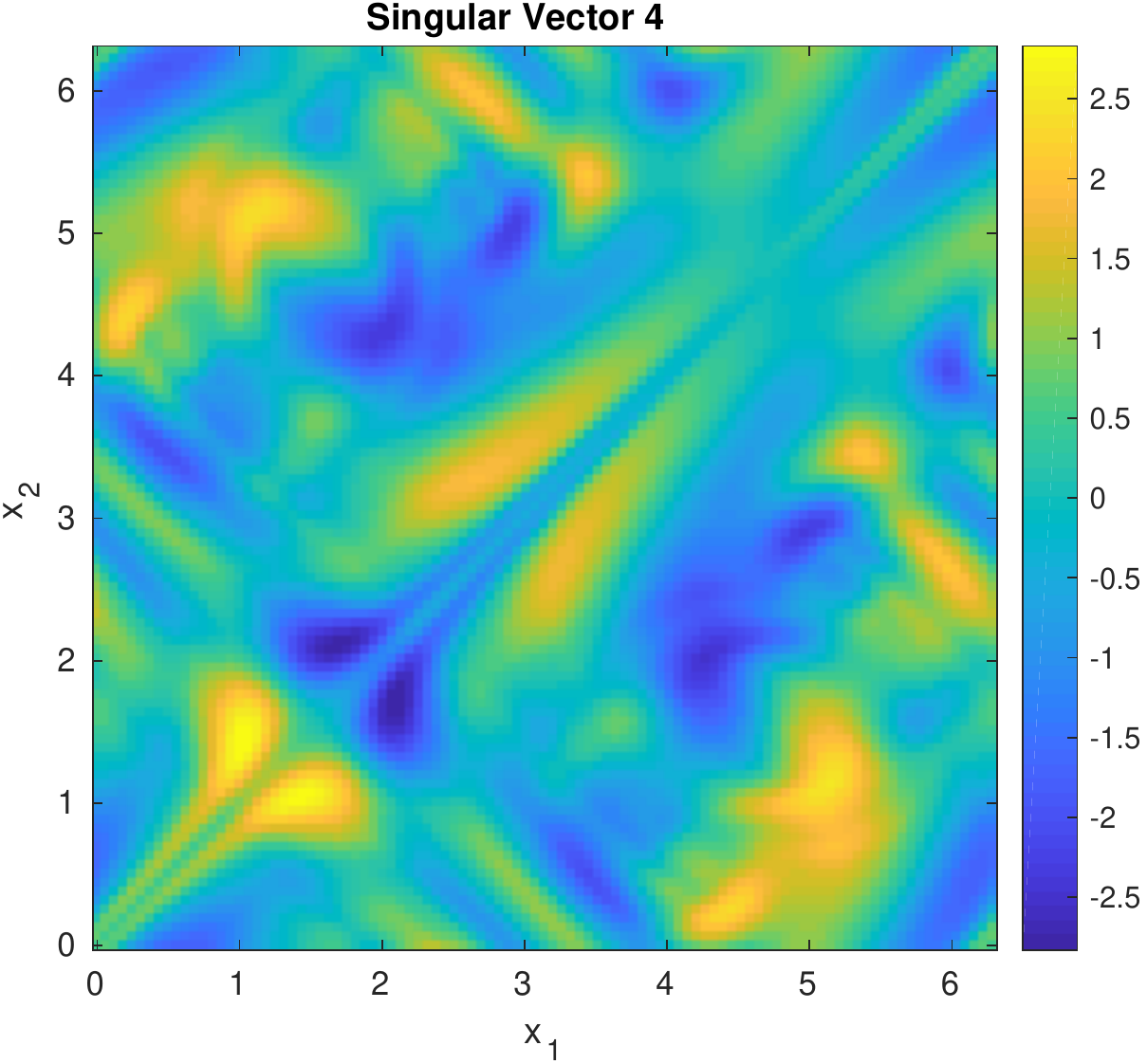}
\includegraphics[width=\plotwidth]{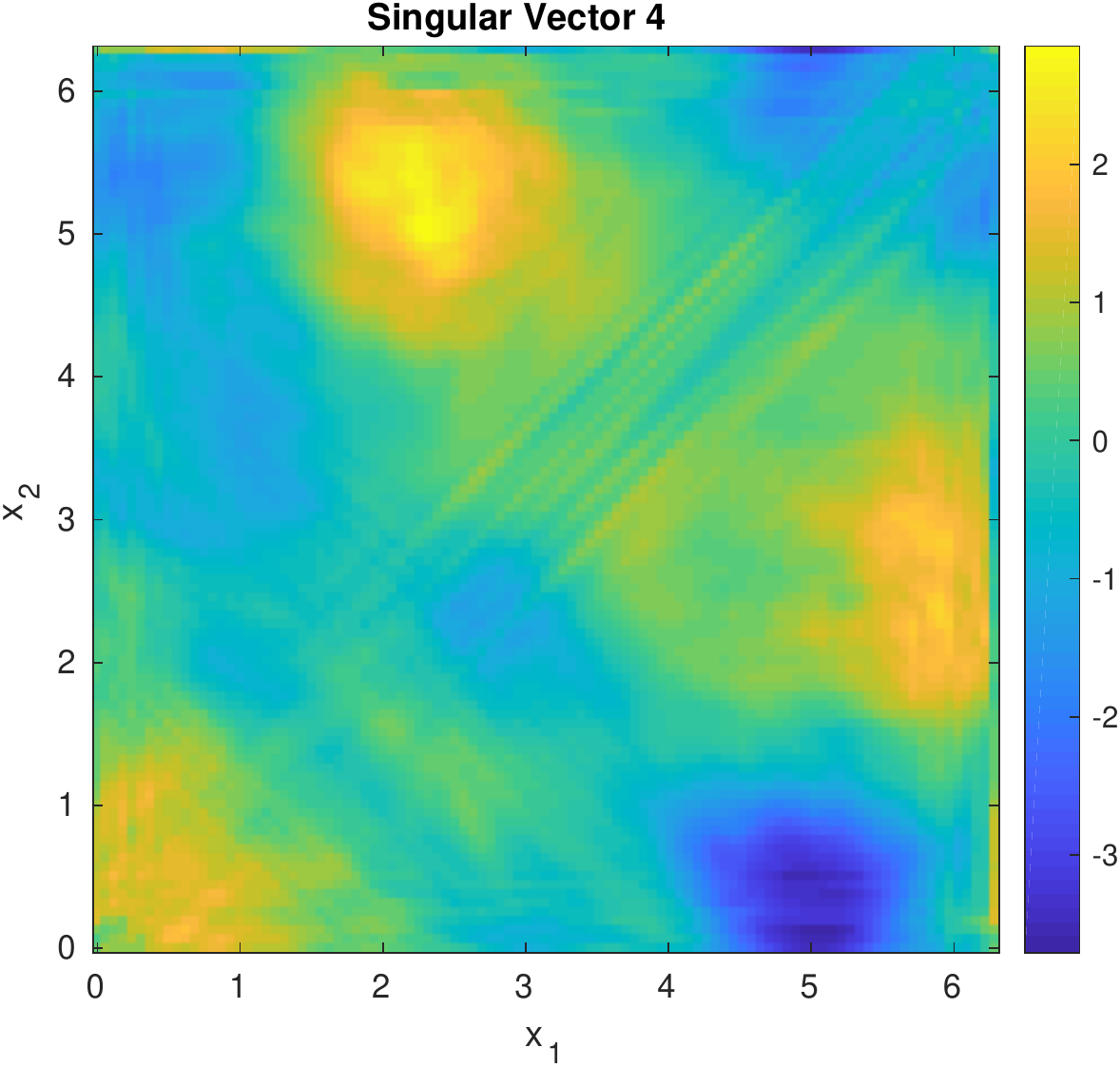}\\
\includegraphics[width=\plotwidth]{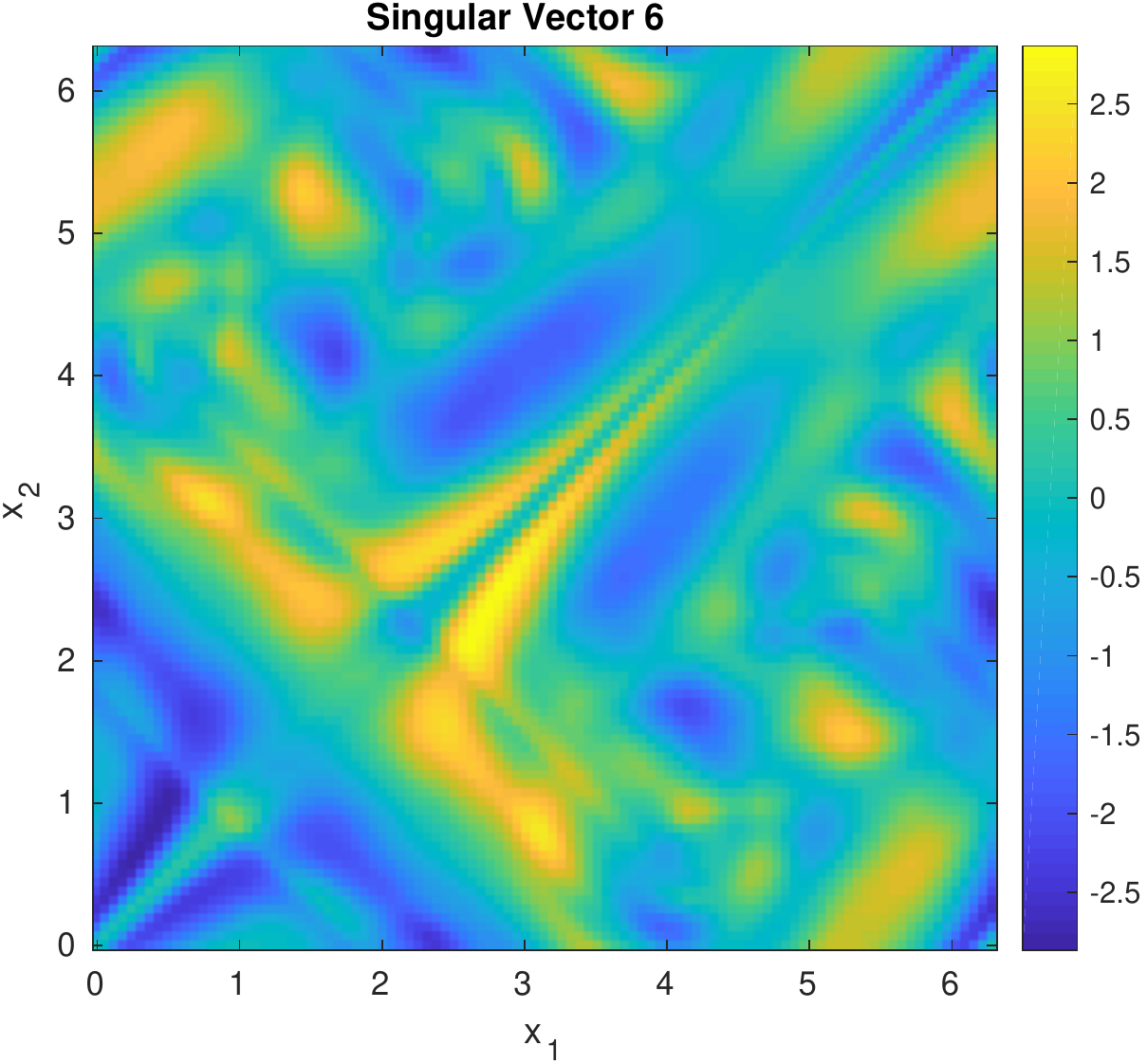}
\includegraphics[width=\plotwidth]{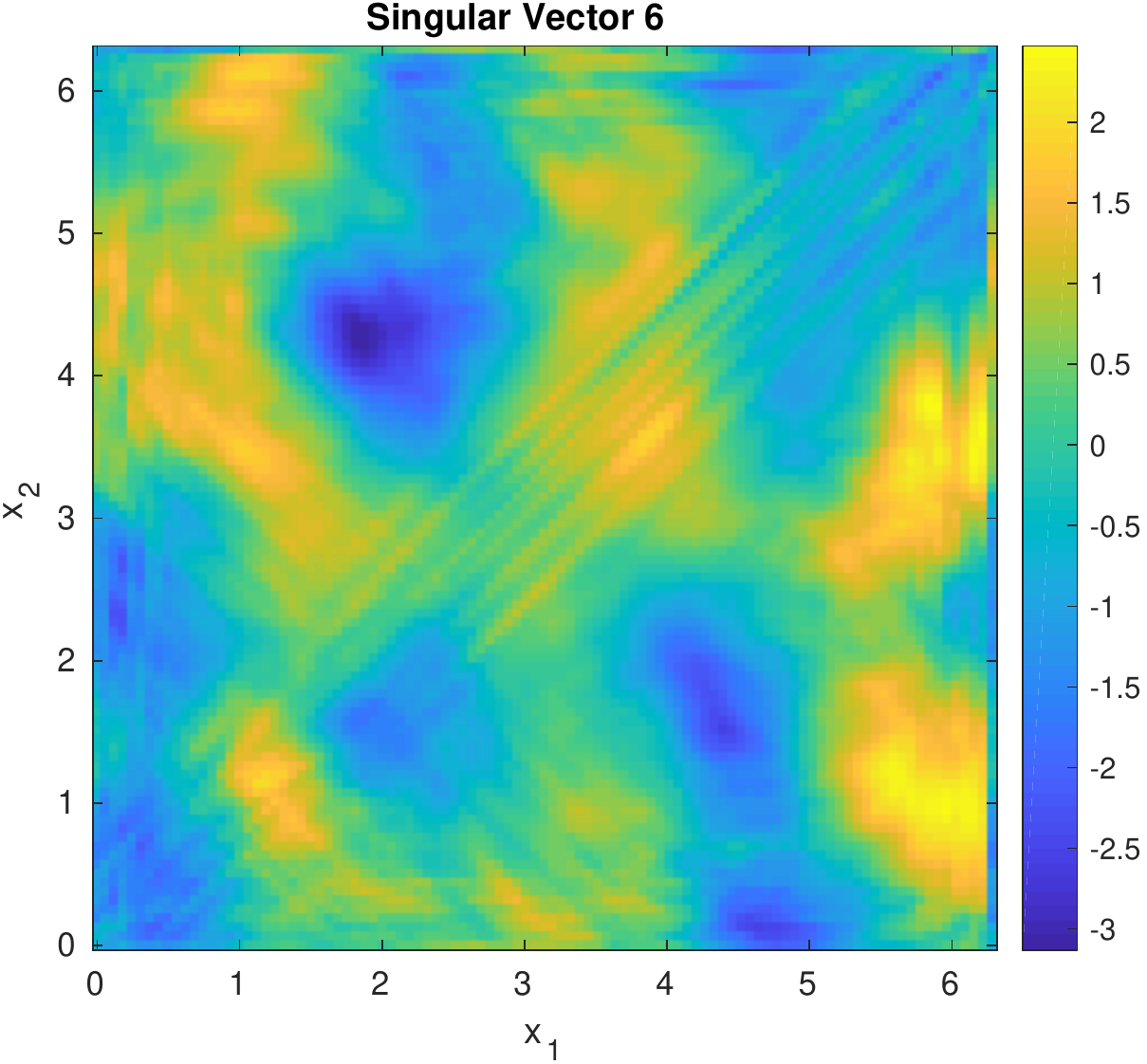}
\end{tabular}
\medskip

\caption{Singular vectors of $W^2$ (left col.),  $\mathcal{F}\left(G\left(\mathcal{F}^{-1}\left(W^2\right)\right)\right)$ (right col.).}\label{srgansv}
\end{figure}

\begin{table}[ht]
    \centering
    \begin{tabular}{|c|c|c|c|c|c|c|}\hline
    \multicolumn{2}{|c|}{$D_0$} & $5\times10^{-4}$ & $4\times10^{-4}$ & $3\times10^{-4}$ & $2\times10^{-4}$ & $10^{-4}$\\\hline
    \multicolumn{2}{|c|}{$\hat{D}^E_{11,60}$} & $1.3847$ & $1.3940$ & $1.4105$ & $1.4395$ & $1.4951$\\\hline
    \multirow{2}{*}{$\hat{D}^{E,a}_{11,50}$} & SVD & $1.5258$ & $1.5597$ & $1.5969$ & $1.6381$ & $1.6854$\\\cline{2-7}
    & SRGAN & $1.2429$ & $1.2663$ & $1.3056$ & $1.3786$ & $1.5293$\\\hline
    \multirow{2}{*}{relative error} & SVD & $10.2\%$ & $11.9\%$ & $13.2\%$ & $13.8\%$ & $12.7\%$\\\cline{2-7}
    & SRGAN & $10.2\%$ & $\boldsymbol{9.2\%}$ & $\boldsymbol{7.4\%}$ & $\boldsymbol{4.2\%}$ & $\boldsymbol{2.3\%}$\\\hline
    \end{tabular}
    \vspace{.1 in}
    
    \caption{Comparison of $\hat{D}^{E,a}_{11,N}$ for flow \eqref{flowform_1} with $D_0^1 = 10^{-2}$, $D_0^2 = 10^{-3}$.}
    \label{srgan1e-2}
\end{table}
When $D_0^1$ is closer to $D_0^2$, SRGAN assisted SVD may have even better predictions at smaller $D_0$. In Table \ref{srgan5e-3}, $D_0^1 = 5\times10^{-3}$, $D_0^2 = 10^{-3}$ and $N = 50$ and we predict the $\hat{D}^E_{11,60}$ at $D_0 = 3\times10^{-4}$, $2\times10^{-4}$ and $10^{-4}$. For $D_0^1 = 5\times10^{-3}$, $D_0^2 = 10^{-3}$, singular vectors of $\mathcal{F}\left(G\left(\mathcal{F}^{-1}\left(W^2\right)\right)\right)$ also have thinner structures than that of $W^2$, as shown in right column and left column of Figure \ref{srgansv} respectively. Table \ref{srgan1e-3} summarizes predictions for $D_0 = 2\times10^{-5}$ and $10^{-5}$ from $D_0^1 = 10^{-3}$, $D_0^2 = 10^{-4}$ and $N = 60$.
\begin{table}[ht]
    \centering
    \begin{tabular}{|c|c|c|c|c|}\hline
    \multicolumn{2}{|c|}{$D_0$} & $3\times10^{-4}$ & $2\times10^{-4}$ & $10^{-4}$\\\hline
    \multicolumn{2}{|c|}{$\hat{D}^E_{11,60}$} & $1.4105$ & $1.4395$ & $1.4951$\\\hline
    \multirow{2}{*}{$\hat{D}^{E,a}_{11,50}$} & SVD & $1.5969$ & $1.6381$ & $1.6854$\\\cline{2-5}
    & SRGAN & $1.3111$ & $1.3862$ & $1.5015$\\\hline
    \multirow{2}{*}{relative error} & SVD & $13.2\%$ & $13.8\%$ & $12.7\%$\\\cline{2-5}
    & SRGAN & $\boldsymbol{7.0\%}$ & $\boldsymbol{3.7\%}$ & $\boldsymbol{0.4\%}$\\\hline
    \end{tabular}
    \vspace{.1 in}
    
    \caption{Comparison of $\hat{D}^{E,a}_{11,N}$ for flow \eqref{flowform_1} with $D_0^1 = 5\times10^{-3}$, $D_0^2 = 10^{-3}$.}
    \label{srgan5e-3}
\end{table}
\begin{table}[ht]
    \centering
    \begin{tabular}{|c|c|c|c|}\hline
    \multicolumn{2}{|c|}{$D_0$} & $2\times10^{-5}$ & $10^{-5}$\\\hline
    \multicolumn{2}{|c|}{$\hat{D}^E_{11,60}$} & $1.6052$ & $1.6301$\\\hline
    \multirow{2}{*}{$\hat{D}^{E,a}_{11,60}$} & SVD & $1.5107$ & $1.5243$\\\cline{2-4}
    & SRGAN & $1.6234$ & $1.7120$ \\\hline
    \multirow{2}{*}{relative error} & SVD & $5.9\%$ & $6.5\%$\\\cline{2-4}
    & SRGAN & $\boldsymbol{1.1\%}$ & $\boldsymbol{5.0\%}$\\\hline
    \end{tabular}
    \medskip
    
    \caption{Comparison of $\hat{D}^{E,a}_{11,N}$ for flow \eqref{flowform_1} with $D_0^1 = 10^{-3}$, $D_0^2 = 10^{-4}$.}
    \label{srgan1e-3}
\end{table}


\section{Conclusions}
Based on snapshots at two molecular diffusivity values, we trained an adapted super-resolution deep neural network (SRGAN) to model the internal layer sharpening effect of 
advection-diffusion equation as a nonlinear mapping. The mapping improves the quality of standard POD basis for low cost computation of residual diffusivity in chaotic flows. 
Though no other DNN model is known to assist POD in our setting, we shall explore how to improve the fidelity of the generated images in the current model. 

\section{Acknowledgements}
The work was supported in part by NSF grants IIS-1632935, and  DMS-1924548.



\clearpage

\begin{thebibliography}{99}

\bibitem{BLP2011} A. Bensoussan, J.-L. Lions, G. Papanicolaou,  
``Asymptotic analysis for periodic structures'', AMS Chelsea Publishing, 2011.

\bibitem{BCVV95} L. Biferale, A. Cristini, M. Vergassola, A. Vulpiani, 
{\em Eddy diffusivities in scalar transport}, Physics Fluids, 7(11), 1995, pp. 2725--2734.

\bibitem{CW91}R. Camassa, S. Wiggins, {\em Chaotic advection in a Rayleigh-B\'enard flow,} 
Physical Review A, 43(2), 1990, pp. 774--797.

\bibitem{CG95} S. Childress, A. Gilbert, 
``Stretch, Twist, Fold: The Fast Dynamo'', Lecture Notes in Physics Monographs, No. 37, 1995, Springer.

\bibitem{FP94} A. Fannjiang, G. Papanicolaou, 
{\em Convection enhanced diffusion for periodic flows,} SIAM J. Applied Math, 54(2), pp. 333-408, 1994.

\bibitem{H03} S. Heinze, 
{\em Diffusion-advection in cellular flows with large Peclet numbers,}
 Archive Rational Mech. Analysis, 168(4), 2003, pp. 329--342.
 
 \bibitem{HL_98} P. Holmes, J. Lumley, G. Berkooz, 
``Turbulence, Coherent Structures,
Dynamical Systems and Symmetry'', 1998, Cambridge University Press.


\bibitem{LTHCCAATTWS2017}
C. Ledig, L. Theis, F. Huszar, A. Cunningham,  A. Acosta, A. Aitken, A. Tejani, J. Totz, Z. Wang, W. Shi, 
{\em Photo-Realistic Single Image Super-Resolution Using a Generative Adversarial Network}, CVPR, pp. 105-114, July 2017.

\bibitem{Lum_67}J. Lumley,
{\em The Structures of Inhomogeneous Turbulent Flows}, Atmospheric Turbulence and Radio Wave Propagation, 
pp. 166-178, 1967. 


\bibitem{LXY_17b}J. Lyu, J. Xin, Y. Yu,  
{\em Computing Residual Diffusivity by Adaptive Basis Learning via Spectral Method,}
Numerical Mathematics: Theory, Methods \& Applications, 10(2), pp. 351-372, 2017. 

\bibitem{MK_99}A. Majda, P. Kramer, 
{\em Simplified Models for Turbulent Diffusion: Theory, Numerical Modelling, 
and Physical Phenomena}, Physics reports, 314(1999), pp. 237-574.


\bibitem{RMC2016}
A. Nasrollahi, L. Metz, S. Chintala, 
{\em Unsupervised Representation Learning with Deep Convolutional Generative Adversarial Networks}, ICLR, 2016.

\bibitem{NR_07} A. Novikov, L. Ryzhik, 
{\em Boundary layers and KPP fronts in a cellular flow}, Arch. Rational Mech. Anal. 184(1), pp. 23-48, 2007.

\bibitem{Quart_14} A. Quarteroni, G. Rozza (eds.),
``Reduced Order Methods for Modeling and Computational Reduction'', 
MS\&A, Vol. 9, Springer, 2014. 

\bibitem{vgg_15}
K. Simonyan and A. Zisserman,
{\em Very deep convolutional networks for large-scale image recognition}, ICLR, 2015.

\bibitem{Sir_87}L. Sirovich,
{\em Turbulence and the dynamics of coherent structures. Part I: Coherent structures},
Quart. App. Math, 45(1987), pp. 561-571.
   
\bibitem{T_21}G. Taylor,
{\em Diffusion by continuous movements}, Proc. London Math. Soc., 2:196--211, 1921.

\bibitem{WXZ_18}Z. Wang, J. Xin, and Z. Zhang, 
{\em Computing effective diffusivity of chaotic and
stochastic 
flows using structure-preserving schemes},  SIAM Journal on Numerical Analysis,
56(4):2322--2344, 2018.

\bibitem{XY_13}J. Xin, Y. Yu,
{\em Sharp asymptotic growth laws of turbulent flame speeds in cellular flows 
by inviscid Hamilton-Jacobi models,}
 Annales de l'Institut Henri Poincar\'e, Analyse Nonlineaire, 30(6), pp. 1049--1068, 2013. 

\bibitem{XY_14a}J. Xin, Y. Yu,
{\em Front Quenching in G-equation Model Induced by Straining of Cellular Flow}, 
Arch. Rational Mechanics \& Analysis, 214(2014), pp. 1-34. 

\bibitem{ZCX_15}P. Zu, L. Chen, J. Xin, 
{\em A Computational Study of Residual KPP Front Speeds in Time-Periodic Cellular Flows 
in the Small Diffusion Limit}, Physica D, Vol. 311-312, pp. 37-44, 2015.


\end{thebibliography}

\end{document}